\documentclass[12pt,epsfig]{article}
\newcommand{\ba}{\begin{array}{c}}
\newcommand{\baz}{\begin{array}{cc}}
\newcommand{\bad}{\begin{array}{ccc}}
\newcommand{\bav}{\begin{array}{cccc}}
\newcommand{\ea}{\end{array}}

\usepackage{amssymb} 
\usepackage{epsfig}
\usepackage{float}
\newcommand{\be}{\begin{equation}}
\newcommand{\ee}{\end{equation}}
\newcommand{\bea}{\begin{eqnarray}}
\newcommand{\eea}{\end{eqnarray}}
\begin{document}

\begin{center}
\bf {The Role of Invariant Functions in Understanding Masses and Mixings}
\end{center}

\begin{center}
C. Jarlskog 
\end{center}

\begin{center}
{\em Division of Mathematical Physics\\
LTH, Lund University\\
Box 118, S-22100 Lund, Sweden}
\end{center}

\begin{abstract}

One of the central questions in theoretical particle physics, since
already several decades, 
has been that of "masses and mixings of the 
quarks. With the entry of neutrino oscillations into the field, 
the issue of lepton masses has added a new dimension to the
problem. In the literature one finds many models 
of quark and lepton mass matrices. However, the mass and mixing
problems remain unsolved. 

In this talk I will present my own contributions to this field,
however as you may expect without offering a solution. I encourage
you to go ahead and think about the problem
but be aware of the pitfalls.

\end{abstract}

\section{Introduction}

Understanding the pattern of masses and mixings of the
fundamental fermions is generally considered to be extremely
important. In the literature one finds several methods of approaching
this problem. Let me consider first the quark sector and mention
only three approaches.
 
The first one could be called the "brute force method". 
One makes ad hoc assumptions
about the quark mass matrices, diagonalises them and determines
the masses and mixings. An ambitious approach within this
category is to "scan" all possible mass matrices. 
Actually, assuming that there are three families, we know the quark mass matrices 
from experiments (though not accurately) since about twenty years already! 
Why then scan and look for something
that you already have at hand? The reason, I believe, is that
one hopes to find an "elegant" solution which may
lead to a deeper insight and understanding.

A second method is to go beyond the standard model. After all,
within the Standard Model, the masses and mixings are arbitrary
parameters to be determined by experiments. But, for example, in grand unified
models one may make predictions for ratios of masses. However,
the situation is not yet satisfactory. 

A third method is to try to detect numerical relations
among either masses or masses and mixings or the
mixing angles themselves, hoping that 
these may indicate a profitable direction for further work. One recent example
is the so-called quark-lepton complementarity relations which have recently
enjoyed some popularity.

\section{Why invariant functions}

In mid 1980's a question which was being asked was
how does one define maximal CP violation in the Standard Model
and is CP maximally violated as is parity? 
As is often the case, there were conflicting opinions 
on what was meant by maximal CP violation.
Some authors were advocating that CP
is maximally violated if the
CP phase in the quark mixing matrix is 90 degrees.
However, such a definition did not make sense because
there is no such unique CP phase. This phase is convention dependent,
your CP phase is in general
a function of my CP phase and mixing angles. Therefore it was
important to understand the structure of CP violation in the
Standard Model. It turned out that there is indeed a 
convention independent invariant
for CP violation \cite{ceja85} and this gave rise to the concept
of invariant functions of mass matrices \cite{ceja85a}. These
were further studied in \cite{ceja87}.

\section{Definition of invariant functions}

The gauge structure of the Standard Model is defined by 
the product of three {\bf local} unitary groups $SU(3)\times SU(2)\times U(1)$.
However, in the quark sector there are additional global
symmetries, as follows. For $n$ families, 
we have as usual $n$ left-handed doublets
as well as $2n$ right-handed singlets,

\begin{equation}
\left(
\begin{array}{c}
u_j \\ d_j
\end{array}
\right), ~~~u^\prime_j, ~~~~~~d^\prime_j
\end{equation}

\noindent where $j=1-n$. Here the $u_j$ and $d_j$ stand for the left-handed
up-type and down-type quarks respectively while their primed
counterparts represent the right-handed singlets. We can thus form
four vectors

\begin{equation}
Q_u = \left(
\begin{array}{c}
u_1 \\ u_2\\.\\u_n
\end{array}
\right), ~~~Q_d= \left(
\begin{array}{c}
d_1 \\ d_2\\.\\d_n
\end{array}
\right), ~~~
Q^\prime_u = \left(
\begin{array}{c}
u^\prime_1 \\ u^\prime_2\\.\\u^\prime_n
\end{array}
\right), ~~~Q^\prime_d= \left(
\begin{array}{c}
d^\prime_1 \\ d^\prime_2\\.\\d^\prime_n
\end{array}
\right)
\end{equation}

In special relativity we learn that physics does not
depend on the choice of the inertial frame of reference. 
Something similar
happens here where the frames are related
to each other by unitary rotations. The transformations between
these frames correspond to:

\vspace{.2cm}
I. Rotating the right-handed up-type quarks by an arbitrary
global unitary matrix, i.e.,
$ Q^\prime_u \rightarrow X^\prime_u Q^\prime_u. $

\vspace{.2cm}
II. Rotating the right-handed down-type quarks by an arbitrary
global unitary matrix,
$ Q^\prime_d \rightarrow X^\prime_d Q^\prime_d. $
 
\vspace{.2cm}
III. Rotating the left-handed up-type and down-type
quarks simultaneously with the same unitary matrix
$ Q_u  \rightarrow X Q_u, ~~~ Q_d \rightarrow X Q_d. $

\vspace{.2cm}
\noindent The $X$'s above are all arbitrary unitary
matrices.

\vspace{.2cm}
IV. In addition to the above symmetries there is also the freedom of choosing
the phases of the physical quarks.

\vspace{.2cm}

In this article, due to lack of allocated space, I will
only briefly list a number of results that I have obtained, 
some in collaboration with colleagues, 
by making use of the above symmetries. 

The first two symmetries can be used 
to make the quark mass matrices Hermitian
without loss of generality , if one so
wishes. This is useful for
certain applications \cite{frampton85}.

\noindent The symmetry denoted by III gives the
largest class of invariant functions.
These functions appear often in physics. 
Indeed what enters in the Standard Model
is not the mass matrices $M_u$ and $M_d$ for the up-type
and down-type quarks respectively but
\begin{equation}
S_u \equiv M_u M_u^\dagger~,~S_d \equiv M_d M_d^\dagger
\end{equation}
These are frame dependent. Under III they change
\begin{equation}
S_u \rightarrow X S_u X^\dagger, ~~~
S_d \rightarrow X S_d X^\dagger
\end{equation}
For the case
of three families one may construct an
invariant function that automatically keeps track
of the 14 conditions that are required to get
CP violation. This function is given by (\cite{ceja85},
\cite{ceja85a}).
\begin{equation}
det \left[ S_u,S_d \right]  = 2i J~ v(S_u) v(S_d)
\label{detquark}
\end{equation}
\noindent where J is an invariant whose magnitude equals  twice the area of
any of the six by now well-known unitarity triangles \cite{stora88}. The quantities
$v(S_u) $ and $v(S_d) $ are given by 
$v(S_u) = (m^2_u-m^2_c)(m^2_c -m^2_t)(m^2_t -m^2_u)$ and 
$v(S_d)= (m^2_d-m^2_s)(m^2_s -m^2_b)(m^2_b -m^2_d)$
(for a pedagogical discussion see \cite{cpboken}).

Looking into literature, we see 
the determinant in Eq.(\ref{detquark}) 
in essentially every computation involving CP
violation, in all its glory when all the six quarks enter on equal footing
but otherwise in a well-defined truncated form, 
where some factors are missing due to
assumptions made in the calculation \cite{ceja87}.
Examples of the first kind are the renormalization
of the $\theta$-parameter of QCD by the electroweak interactions and 
the calculation of the
baryon asymmetry of the universe in the standard model. An example of
the second kind is the computation of
the electric dipole moment of a quark, say the
down quark. Since in such a computation, the down quark appears in the external legs,
it is tacitly assumed
that we know the identity of this quark, i.e., $m_d \neq m_s$ and $m_d \neq m_b$.
Therefore the factors
$(m^2_d-m^2_s)$ and $(m^2_b -m^2_d)$ are missing but all
the other factors are present.

The above commutator is the simplest in a family of commutators of
functions of mass matrices \cite{ceja85a} which I will not discuss here.
These give us new invariant functions, some of which enter in the computation
of neutrino oscillations \cite{cejanu}. In addition, for example,
the absolute values of the elements of the quark mixing matrix
are measurable quantities and thus invariant functions. These functions
were constructed in \cite{ceja87} (see also \cite{cejaproj}).

\section{Textures and quark-lepton relations}

Theorists love zeros. Generally zeros in the mass matrices
make their diagonalisation and thus life simpler.
However, we have seen that such zeros are not invariants!
If you give me mass matrices with some zeros I may
use symmetry III whereby the zeros could 
evaporate. In the case of leptons in most models one ends up
with an effective three-by-three mass matrix for the
neutrinos. In this case there is no known justification
for making assumptions about how the neutrino mass matrix
should look like in a frame where the charged-lepton mass matrix
is diagonal. This frame is 
highly "fine-tuned". You have to ask yourself
how does the Standard Model find this very special frame
in an infinite class of all possible frames?

My problem with the quark-lepton complementarity relations
(see the talk by Mohapatra where the subject is briefly
discussed) is again that such relations are not invariants.
Mixing angles, for three or more families, are convention
dependent. Your mixing angles and phases are in general
functions of my corresponding quantities. Even the order
in which the Euler-type rotations are introduced is not
preset. Let us take a very simple example. 
As mentioned before, in most
models the leptonic mixing matrix appears essentially in
the same form as the quark mixing matrix. In a simplified
notation we have 
\begin{equation}
(e, \mu , \tau ) U \left( \begin{array}{c} \nu_1 \\ \nu_2 \\
\nu_3 \end{array} \right)
\end{equation}
Suppose that we decide to write the mixing matrix $U$ in two 
different but equivalent forms
\begin{equation}
V= R_{23} R_{13} R_{12}, 
~~~ V^\prime= R^\prime_{12} R^\prime_{13} R^\prime_{23}
\end{equation}
where $R_{jk}$ denotes a rotation in the $jk$-plane by
the angle $\theta_{jk}$ while its primed counterpart denotes
the same rotation by the angle $\theta^\prime_{jk}$. 
In this example the second matrix is obtained from the first
one by transposition and redefinition of the angles, 
$ V^\prime = \tilde{V} (\theta_{jk} \rightarrow -\theta^\prime_{jk}).$ 

Now compare the two limits $\theta_{13} \rightarrow 0$ and $\theta^\prime_{13}$.
For nondegenerate neutrinos they are physically inequivalent. 
In one case the electron couples only to two 
physical neutrinos while the muon and
the tau couple to all three. In the other case it is the tau that couples
only to two neutrinos. This simple example shows that it doesn't make sense
to compare mixing angles as these are ill-defined. The problems become
more serious when one adds mixing angles from the quark and lepton sectors.
We don't even know whether we should add or subtract, what are the relative
phases, etc. Some of these issues have been discussed in \cite{cejaqlc}.

\section{Outlook}

The issue of masses and mixings remains an unsolved problem
that deserves our attention. However, we should always keep
in mind that important results can't be frame dependent.
So, if you have an important message to transmit, you should
be able to formulate it in an invariant form.
Unfortunately, I have no space to describe some work done
along this line \cite{harrison}, nor some
work that I have done on parameterisation of unitary matrices
\cite{cejarecur} which may be of interest to you.

\section{Acknowledgements}
I wish to thank the Organisers of the SNOW 2006, specially Tommy Ohlsson, 
for a wonderful meeting where this talk was presented.

\end{document}